\documentclass[twocolumn,showpacs,preprintnumbers,amsmath,amssymb]{revtex4}

\usepackage{graphicx}
\usepackage{dcolumn}
\usepackage{bm}

\begin{document}

\title{Effective Magnetic Fields in Graphene Superlattices}

\author{Jianmin Sun}
\affiliation{Department of Physics, Indiana University, Bloomington, IN 47405}
\author{H.A. Fertig}
\affiliation{Department of Physics, Indiana University, Bloomington, IN 47405}
\author{L. Brey}
\affiliation{Instituto de Ciencia de Materiales de Madrid
(CSIC), Cantoblanco 28049, Spain}

\date{\today}

\begin{abstract}
We demonstrate that the electronic spectrum of graphene in a one-dimensional
periodic potential will develop a Landau level spectrum when
the potential magnitude varies slowly in space.  The effect is related
to extra Dirac points generated by the potential whose
positions are sensitive to its magnitude.  We develop an effective
theory that exploits a chiral symmetry in the Dirac Hamiltonian description
with a superlattice potential, to show that the low energy theory contains an
effective magnetic field.  Numerical diagonalization of the Dirac equation
confirms the presence of Landau levels.  Possible consequences
for transport are discussed.
\end{abstract}

\pacs{72.80.Vp,73.21.Cd,73.22.Pr} \maketitle

Graphene is a honeycomb lattice of carbon atoms with a host of remarkable
electronic properties \cite{Castro_Neto_RMP,dassarma_2010}.  Many of
these are a result of the low energy states, which are
governed by a massless Dirac equation.  When undoped the Fermi
surface of the system is a set of discrete points, with resulting
properties that compromise between metallic and insulating behavior.
For example, an ideal graphene system can conduct diffusively
even in the absence of impurities to scatter the electrons
\cite{Tworzydlo_2006,Brey_2007c}.  The possibility of a universal
conductivity even in the presence of impurities remains an
active subject of discussion
\cite{Castro_Neto_RMP,dassarma_2010,Novoselov_2004,Novoselov_2005,Zhang_2005}.
The ``relativistic'' nature of the electronic spectrum introduces
Klein paradox physics \cite{katsnelson_2006,Castro_Neto_RMP}
which may be evident in $p-n$ junctions and
related structures \cite{Stander_2009,Young_2009,Cheianov_2007,Zhang_2008,Arovas_2010}.
Beyond this, graphene exhibits signatures of ``effective time-reversal
symmetry breaking," effects most naturally described in terms
of effective magnetic flux, due to a Berry's phase intrinsic
to graphene \cite{Beenakker_2008,Luo_2009,Rycerz_2007},
or to strain in the lattice system \cite{Guinea_2009,Guinea_2010b,Vozmediano_2010},
even in the absence of a real applied magnetic field.

In this paper we discuss a completely different way to induce
an effective magnetic field in graphene, of sufficient uniformity
that Landau levels in principle should be apparent in the
electronic spectrum.  This involves immersing the graphene
system in a unidirectional periodic ({\it i.e.}, superlattice) potential,
for example by subjecting a sample with periodic ripples \cite{Lau_2009}
to a perpendicular electric field.  Recently it has been shown
\cite{Brey_2009,Park_2009} that such a potential can induce new
Dirac points from the original one, whose positions are determined
by the amplitude $V_0$ of the periodic potential, and by its period.
We demonstrate that by promoting $V_0$ to a slowly varying function of position,
the effective theory for states near an induced Dirac point contains
a gauge field, such that there are states in the spectrum corresponding
to Landau levels.

\begin{figure}
  \includegraphics[clip,width=8.5cm]{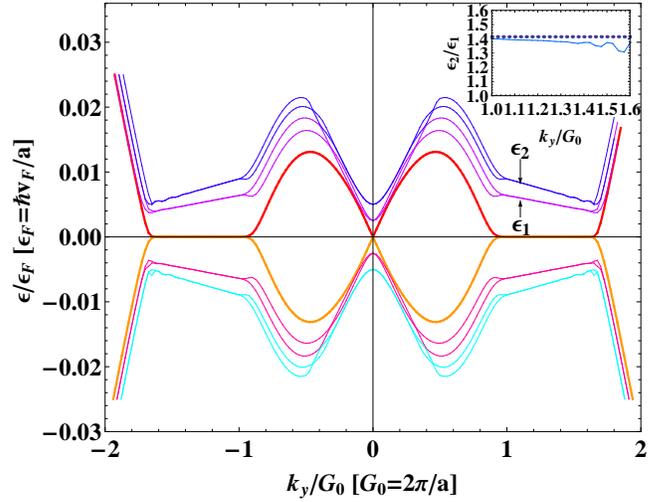}
  \caption{($Color$ $online$)
  Energy bands for Dirac equation in a modulated superlattice potential.
  Inset: Ratio of first to second excited state energies near $k_ya/2\pi=1.3$,
  demonstrating the $\sqrt{2}$ ratio (dotted line) expected of Landau levels.}
   \label{Figure1}
\end{figure}

The possibility of generating such a description turns out to be intimately
related to a chiral symmetry which is present in the Dirac equation even with a
periodic potential, as we show below.  The symmetry requires that any
state with energy $\varepsilon$  have a chiral partner at energy $-\varepsilon$.
These
are degenerate for $\varepsilon=0$.  Eigenstates near zero energy
can be approximately constructed from these chiral partners, and they
have a number of properties guaranteeing that the resulting effective
theory has the form of a Dirac equation with a vector potential,
representing a magnetic field with strength proportional to the rate
at which the position of the induced Dirac point varies with position.
We demonstrate the effect explicitly by numerically solving the
Dirac equation with a potential $V_0(x)\cos G_0x$.  Results for the energy spectrum
are illustrated in Fig. \ref{Figure1}.  Within this
spectrum we find
level spacings that are essentially identical to those of Landau levels (see inset
of Fig. \ref{Figure1}),
and wavefunctions contained in envelopes whose center positions -- guiding
centers -- are fixed by a wavevector, and have the forms of harmonic
oscillator states.  We find that states that straddle
regions in which $dV_0/dx$ changes sign are strongly dispersive, acting
much like edge states of graphene ribbons in a magnetic field \cite{Brey_2006b}.
The presence of such edge states should have very interesting consequences
for transport in this system.

\textit{Superlattice Potential and Chiral Symmetry} -- Low energy states of graphene
in a single valley may be modeled by a Dirac equation \cite{Ando_2005}.
When an external potential is added, the eigenvalue equation to be
solved has the form $H\vec{\psi}=\varepsilon\vec{\psi}$, with
$H=v_F (\sigma_x p_x+\sigma_y p_y) + V_0v(x) {\cal I}$,
where $\vec{\psi}$ is a two-component spinor whose entries are
the wavefunction amplitudes for the two sublattices, $v_F$ is the
speed of electrons with momenta near the Dirac point,
$\sigma_{x,y}$ are Pauli matrices, $p_{x,y}=-i\hbar\partial_{x,y}$,
$V_0$ is the scale of an external potential with functional form $v(x)$,
and ${\cal I}$ is a unit $2\times 2$ matrix.  For $V_0=0$, $H$ is
known to have a chiral symmetry, which can be expressed as $\{\sigma_z,H\}=0$,
with $\sigma_z$ the third Pauli matrix.  This anticommutator relation
implies that for any eigenstate $\vec{\psi}$ with energy $\varepsilon$,
there is a particle-hole partner $\sigma_z \vec{\psi}$ which is also
an eigenstate of $H$, with energy $-\varepsilon$.  Moreover one may use
this property to show that the state at zero energy is doubly degenerate.

When there is a non-vanishing periodic superlattice potential
satisfying $v(x+a/2)=-v(x)$ with energy scale $V_0$, it was shown
recently \cite{Brey_2009,Park_2009}
that for sufficiently large $V_0 a$ one obtains new Dirac cones
in the spectrum.  These appear as pairs which emerge from ${\bf k}=0$ along
the $k_y$ axis, one moving up and the other down as $V_0 a$ increases,
so that there are new zero energy states at ${\bf k}=(0,\pm k_y^*)$.
Further increasing $V_0a$ generates yet more pairs of new Dirac cones.
For simplicity, in what follows we will only consider the situation
where a single new pair has emerged.

The Hamiltonian with this type of periodic potential also has chiral symmetry,
although the associated operator is slightly more complicated.  Defining a
shift operation $S\vec{\psi}(x,y) = \vec{\psi}(x+a/2,y)$, it takes the form
$T=\sigma_zS$; one may easily demonstrate $\{H,T\}=0$ even in the presence
of the periodic potential.  Because there is a chiral operator, one may show
that a zero energy state $\vec{\phi}_0$ at $(0,k^*_y)$ has a
chiral partner $T\vec{\phi}_0$ which is also at zero energy at the  {\it same}
wavevector.  For what follows it is convenient to define eigenstates
of $T$, $\vec{\phi}_{\pm}=[\vec{\phi}_0 \pm T\vec{\phi}_0]/{\cal N}_{\pm}$,
where ${\cal N}_{\pm}$ are normalization integrals.  These states have
a number of interesting properties which will be useful to us:
\begin{eqnarray}
<\phi_+|\phi_-> &=&0, \label{prop1}\\
<\phi_{\pm}|\sigma_{x,y}|\phi_{\pm}> &=& 0, \label{prop2} \\
<\phi_+|\sigma_{x}|\phi_-> &=& iu_x, \label{prop3} \\
<\phi_+|\sigma_{y}|\phi_-> &=& u_y, \label{prop4}
\end{eqnarray}
where $u_{x,y}$ are purely real.  Eq. \ref{prop1} is easily shown by observing that
$\vec{\phi}_{\pm}$ are eigenstates of $T$ with different eigenvalues.  Eq. \ref{prop2}
follows from $\{\sigma_{x,y},\sigma_z\}=0$ and
$<\phi_{\pm}|T^2|\phi_{\pm}>=1$.  Eqs. \ref{prop3}
and \ref{prop4}
follow from a specific form which $\vec{\phi}_0$ may take,
$\vec{\phi}_0({\bf r})=\tilde{\phi}(x)[i\sin\theta(x),\cos\theta(x)]^{\dag}e^{ik^*_yy}$,
where $\tilde{\phi}$ and $\theta$ are real periodic functions \cite{Brey_2009}.
The quantities $u_{x,y}$ may be understood as the velocities
of the electron states in the $(x,y)$ directions of a Dirac cone, in units
of $v_F$, as we shall see below.

\textit{Modulated Superlattice} -- We now consider the situation $V_0 \rightarrow V_0(x)$,
where $V_0$ varies slowly on the scale of the superlattice parameter $a$.
The eigenfunctions generated from $\vec{\phi}_0$ have an extra spatial dependence
from their implicit dependence on $V_0$:
$\vec{\phi}_0(x,y;V_0) \rightarrow \vec{\phi}_0[x,y;V_0(x)]$.  In what follows it will
be convenient to consider functions where the explicit spatial dependence
and that inherited from $V_0(x)$ are independent, so that we promote the
wavefunctions to functions of two variables:
$\vec{\phi}_0[{\bf r^{\prime}};V_0({\bf r})] \rightarrow \vec{\phi}_0({\bf r^{\prime}},{\bf r})$.
Note that in this construction, if ${\bf r}$ is held constant then
$\vec{\phi}_0(x^{\prime}+a,y^{\prime};{\bf r})=\vec{\phi}_0(x^{\prime},y^{\prime};{\bf r})$.
We can then define matrix elements of the form
$$
(\phi_1|A|\phi_2)_{\bf r} \equiv
\int_{x-a/2}^{x+a/2} dx^{\prime} \int_{-\infty}^{\infty} dy^{\prime}
g(y-y^{\prime})\vec{\phi}_1({\bf r^{\prime}},{\bf r})^{\dag} A
\vec{\phi}_2({\bf r^{\prime}},{\bf r}),
$$
where $g(y-y^{\prime})$ is a peaked function centered at 0 and of width much greater than
$2\pi/k^*_y$ \cite{Ando_2005} which integrates to 1,
and $A$ is an operator.  Identifying
$\vec{\phi}_{\pm} \equiv \vec{\phi}_{\pm}({\bf r^{\prime}};V_0(x))$,
Eqs. \ref{prop1}-\ref{prop4} remain true if we replace
the standard matrix elements with those above.  We now consider a generalized
eigenvalue problem
$H({\bf r}^{\prime},{\bf r}) \vec{\psi}({\vec {\bf r}}^{\prime},{\bf r})
=
\varepsilon \vec{\psi}({\vec {\bf r}}^{\prime},{\bf r})$, where
$$
H({\bf r}^{\prime},{\bf r}) =
v_F [-i\sigma_x(\partial_{x'}+\partial_x) -i\sigma_y(\partial_{y'}+\partial_y) ]
+V_0(x)v(x^{\prime}){\cal I},
$$
and we have set $\hbar=1$.
In the limit ${\bf r}^{\prime} \rightarrow {\bf r}$, this becomes the
eigenvalue equation one needs to solve for the physical Hamiltonian; the
advantage of the generalized problem is that it retains the chiral
symmetry $\{H,T\}=0$, with the shift operation ($S$) contained
in $T$ acting only $x^{\prime}$.  To approximately
solve this problem, we consider solutions of the form \cite{Ando_2005}
$$
\vec{\psi}({\bf r}^{\prime},{\bf r}) =
\psi_+({\bf r})\vec{\phi}_+({\bf r}^{\prime},{\bf r})
+
\psi_-({\bf r})\vec{\phi}_-({\bf r}^{\prime},{\bf r}).
$$

\begin{figure}
  \includegraphics[clip,width=8cm]{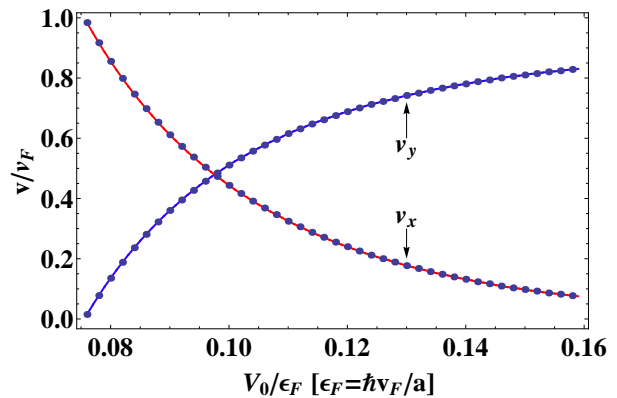}
  \caption{($Color$ $online$) Comparison of velocities $v_{x,y}$, as
  computed from Eqs. \ref{prop3} and \ref{prop4} (solid lines), and from the numerically
  generated band structure in a superlattice potential (dots), around an induced
  Dirac point, as a function of $V_0$.
 }
   \label{vx+vy}
\end{figure}

Acting on this wavefunction with $H$, and then forming matrix elements
$(\phi_{\pm}|H-\varepsilon|\psi)_{\bf r}$, we seek choices of the
coefficients ${\psi}_{\pm}$ for which these matrix elements vanish.
After some algebra \cite{suppl}, with the use of Eqs. \ref{prop1} - \ref{prop4},
one finds that they must obey the equations
\begin{eqnarray}
\left(
\begin{array}{cc}
0 & v_x\partial_x - i v_y\partial_y + a_{+-}({\bf r}) \\
-v_x\partial_x - i v_y\partial_y + a_{-+}({\bf r}) &  0
\end{array}
\right)
\left(
\begin{array}{c}
\psi_+ \\
\psi_-
\end{array}
\right) \nonumber
\\
\,\,\,\,\,\,
=
\varepsilon
\left(
\begin{array}{c}
\psi_+ \\
\psi_-
\end{array}
\right).\quad\quad\quad\quad\quad\quad\quad\quad
\label{almost_dirac}
\end{eqnarray}
Here $v_{x,y}(x)=v_F u_{x,y}[V_0(x)]$, and
$$
a_{\pm,\mp}({\bf r})=
-i v_F(\phi_{\pm}|\vec{\sigma}\cdot\vec{\nabla} | \phi_{\mp} )_{\bf r}.
$$
It is important to note because the gradient operator in $a_{\pm,\mp}$
acts on ${\bf r}$ and not ${\bf r^{\prime}}$, $a_{\pm} \ne a_{\mp}^*$.
Thus the matrix appearing in Eq. \ref{almost_dirac} is not Hermitian.
However, one may show that
${\rm Im}a_{\pm}=-{\rm Im}a_{\pm}=v_x y \partial_x k_y^*
\equiv v_x A_x$
is purely antisymmetric, while the real antisymmetric component of $a$,
$a_A^R \equiv {\rm Re}(a_{\pm}-a_{\mp})/2$,
can
be removed by a similarity transformation. Rewriting Eq. \ref{almost_dirac} in terms of
$\tilde{\psi}_{\pm}=\exp[\int^x d\tilde{x} \, a_A^R(\tilde{x})
v_y(\tilde{x})/v_x(\tilde{x})]  \psi_{\pm}$,   we obtain
a Hermitian eigenvalue equation $\tilde{H}\tilde{\psi}=\varepsilon\tilde{\psi}$, with
\begin{eqnarray}
\tilde{H}=
\sigma_y v_x[-i\partial_x - A_x] + \sigma_x v_y[-i \partial_y -A_y],
\label{dirac_with_field}
\end{eqnarray}
where $A_y=-{\rm Re}(a_{\pm}+a_{\mp})/2v_y$.
Eq. \ref{dirac_with_field} has the form of
a Dirac Hamiltonian with a vector potential, albeit with the roles of the $x$ and $y$
momenta interchanged, and the effective velocities $v_{x,y}(x)$ being direction
and position dependent.  Several remarks are in order.
(i) For $V_0$ constant, $v_{x,y}$ lose their position
dependence, and we expect them to give the velocities of the induced
Dirac points for an unmodulated superlattice potential.  Fig. \ref{vx+vy}
illustrates by comparison with velocities found numerically around
an induced Dirac point of a graphene superlattice \cite{Brey_2009,Park_2009}
that this is indeed the case.  (ii) Chiral symmetry plays a central role
in giving Eq. \ref{dirac_with_field} its Dirac form, through the properties
of the $\vec{\phi}_{\pm}$ states expressed in Eqs. \ref{prop1}-\ref{prop4}.
In particular they guarantee that diagonal elements do not appear in the
Hamiltonian, and that the off-diagonal terms have a form that can naturally be
grouped together into canonical momenta with vector potential contributions.
(iii) The effective field represented by the vector potential is
$B_{eff}=\partial_xA_y-\partial_yA_x$.  In the limit of a slowly
varying modulation ($|a(\partial_xV_0)/V_0| \ll 1$), the first term is
second order in derivatives and is negligibly small, so that the effective
field is $B_{eff}=-\partial_x k^*_y \equiv -(\partial_{V_0}k_y^*[V_0])
(\partial_x V_0(x))$.  We see that the presence
of an induced magnetic field is intimately connected with the
fact that the position of the Dirac point in the Brillouin zone can
be changed via an external parameter.  (iv) With a gauge transformation,
$\tilde{H}$ can be written in a form for which there is no explicit
$y$ dependence, so that $k_y$ is a good quantum number, as we expect
because the system is translationally invariant along the $\hat{y}$  direction.
If we define a magnetic length $1/\ell_0^2 \equiv \partial_x k_y^*$
then $X=k_y\ell_0^2$ becomes a guiding center coordinate around which
the wavefunction is centered.  Approximating $v_{x,y}(x) \rightarrow v_{x,y}(X)$,
$\tilde{H}$ becomes a standard Hamiltonian for Dirac electrons in a
magnetic field; the approximation is valid provided $\ell_0$ is sufficiently
small.  In this case one expects eigenenergies of the form
$\varepsilon_n(X) = sgn(n) \sqrt{2v_xv_y|n|B_{eff}}$ with $n$ an integer.
Although the values of $\varepsilon_n(X)$ depend on $k_y$ through
$v_xv_yB_{eff}$,
neighboring energy levels $n_1$, $n_2$ at fixed $k_y$ should obey
$\varepsilon_{n_1}/\varepsilon_{n_2}=\sqrt{n_1/n_2}$.  This is an
unambiguous signature of Landau levels, which we next demonstrate numerically
to be present in the energy spectrum of this system.

\begin{figure}
  \includegraphics[clip,width=8.5cm]{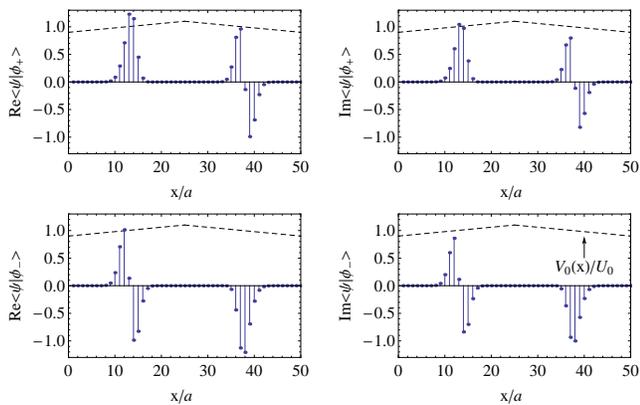}
  \caption{($Color$ $online$) Wavefunctions for $k_ya/2\pi=1.25$,
for a potential modulation $V_0(x)$ of the form illustrated by the dashed line.
 }
   \label{wavefunctions}
\end{figure}

\textit{Numerical Studies} -- Returning to the original Hamiltonian,
$H=v_F (\sigma_x p_x+\sigma_y p_y) + V_0(x)v(x) {\cal I}$,
as a concrete example we take $V_0(x)$ to have a sawtooth
shape (illustrated as a dashed line in Fig. \ref{wavefunctions}), acting
as an envelop for a pure cosine potential $v(x)=\cos{G_0x}$, with $G_0=2\pi/a$.
The period of $V_0$ is chosen to be $L_0=50a$, and its minimum/maximum values
are $(1\pm 0.1)U_0$, with $U_0=0.12\epsilon_F$ and $\epsilon_F=\hbar v_F/a$.
We expand $H$ in plane
wave states, for fixed values of $k_x$ (the Bloch wavevector) and $k_y$, and
diagonalize the resulting matrix.
A typical example of a few of the eigenvalues as a function of $k_y$, for $k_x=0$,
are illustrated in Fig. \ref{Figure1}.
Near $k_y=0$ the Dirac point which
would also be present for $V_0=0$ is apparent, but at larger $|k_y|$ one
finds bands that vary nearly linearly with $k_y$, in the vicinity of the
emergent Dirac points one would find if $V_0(x)$ were replaced by its spatial
average in the potential \cite{Brey_2009,Park_2009}.  These are the Landau
levels generated by the potential modulation.  The zero energy states one
expects of Dirac Landau bands are very apparent.
The inset of Fig. \ref{Figure1}
shows the ratio of energies for the two lowest positive energy states; for
Landau levels these should scale as $\sqrt{n_2/n_1}=\sqrt{2}$, which is indeed
the case as illustrated in the inset.  The levels shown are each very
nearly doubly degenerate.  This is because our chosen $V_0(x)$ satisfies
$dV_0(x)/dx=-dV_0(x+L_0/2)/dx$, so
that the effective magnetic field has regions of equal magnitude but opposite
directions.   We find that the spectra in the vicinity of the Landau
levels are quantitatively nearly identical for any choice of $k_x$.

Fig. \ref{wavefunctions} illustrates wavefunction overlaps
with $\vec{\phi}_{\pm}$ for the lowest
positive energy states for a single choice of $k_y$.
Their similarity to
a lowest harmonic oscillator state for one of these, and to
a first excited excited state for the other, is apparent.
This is precisely what one expects for eigenstates of the Dirac
equation in a uniform magnetic field \cite{Castro_Neto_RMP}.
Note that the wavefunction has support in both the positive
and negative effective field regions, but the roles of $\vec{\phi}_+$
and $\vec{\phi}_-$ which multiply the lower or higher
harmonic oscillator states are reversed.  This reflects the fact
that Landau level raising and lowering operators interchange roles
when the field sign is reversed.

It is interesting to note that the degeneracy of the positive and negative
field states breaks down when $k_y$ is chosen so that the wavefunction
is significantly different than zero at the cusps of $V_0(x)$.  In this
situation the energy rises quickly as a function of $k_y$, indicating a large
current.  These behave much as edge states in a quantum Hall system, and
should have interesting consequences for transport.  In particular we
expect that $\sigma_{yy}$, the conductance along the $\hat{y}$ direction,
should have steps as a function of Fermi energy $E_F$, occurring at energies
approximately reflecting the $\sqrt{nB_{eff}}$ behavior of the Landau levels.
An observation of this effect would constitute evidence of the effective
magnetic field present in this system.
It is also interesting to note that left- and right-moving states are spatially
separated, existing separately at the upward and downward cusps of $V_0(x)$.
We thus expect that backscattering due to disorder will be somewhat suppressed, although
scattering into a
time-reversed partner from the opposite valley will be present if disorder
at short wavelengths is significant in the system.
Studies of this and related transport
phenomena are currently underway.

In summary, we have demonstrated that an effective magnetic field can
be generated in a graphene superlattice by a slow modulation of the superlattice
potential amplitude.  A description in terms of a gauge field can be produced
from zero energy states at a Dirac point of the unmodulated superlattice which
are chiral partners.  These may be used as a basis to find states in the vicinity
of the modulated potential.  For a slowly varying amplitude, it was found that
the effective field is produced by the changing position of the Dirac point
in real space.  Numerical studies of the Dirac equation with this type of
potential verified the presence of Landau levels in the spectrum, and
suggest the possibility of interesting transport phenomena in this system.

{\it Acknowledgements} -- This work was financially supported by the NSF through
Grant No. DMR-0704033, and by MEC-Spain via Grant No. FIS2009-08744.


%
%
%
%
%
%

\widetext
\section*{Effective Magnetic Fields in Graphene Superlattices, Supplementary Information:\\
Derivation of Equation 5}

In this supplement, we provide some details of how Eq. 5 is obtained from
the generalized eigenvalue problem,
$H({\bf r}^{\prime},{\bf r}) \vec{\psi}({\vec {\bf r}}^{\prime},{\bf r})
=
\varepsilon \vec{\psi}({\vec {\bf r}}^{\prime},{\bf r})$, with
\begin{equation}
H({\bf r}^{\prime},{\bf r}) =
\hbar v_F [-i\sigma_x(\partial_{x'}+\partial_x) -i\sigma_y(\partial_{y'}+\partial_y) ]
+V_0(x)v(x^{\prime}){\cal I}.
\label{hamiltonian}
\end{equation}
We consider approximate solutions of the form
$$
\vec{\psi}({\bf r}^{\prime},{\bf r}) =
\psi_+({\bf r})\vec{\phi}_+({\bf r}^{\prime},{\bf r})
+
\psi_-({\bf r})\vec{\phi}_-({\bf r}^{\prime},{\bf r}),
$$
where $\vec{\phi}_{\pm}=[\vec{\phi}_0 \pm T\vec{\phi}_0]/{\cal N}_{\pm}$,
and $H_0\vec{\phi_0}=0$, with
\begin{equation}
H_0({\bf r}^{\prime},{\bf r}) =
\hbar v_F [-i\sigma_x\partial_{x'} -i\sigma_y\partial_{y'} ]
+V_0(x)v(x^{\prime}){\cal I}.
\label{hamiltonian}
\end{equation}
In this context, the operator $T$ is defined as $T\vec{\phi}({\bf r}^{\prime},{\bf r}) \equiv
\sigma_zS\vec{\phi}({\bf r}^{\prime},{\bf r}) \equiv
\sigma_z\vec{\phi}({\bf r}^{\prime}+a\hat{x}/2,{\bf r})$.  Since ${\bf r}$ is
a fixed parameter, the potential in $H_0$ is a periodic function of
$x^{\prime}$, and
$\phi_0$ and $\phi_{\pm}$ are zero energy Bloch states
with respect to ${\bf r}^{\prime}$.  As shown in Refs. 21 and 22, such states
are present at $k_x=0$ and $k_y=k_y^*$, with $k_y^*$ depending on $V_0(x)$.
Applying Eq. \ref{hamiltonian} to $\vec{\psi}$, multiplying by $g(y-y^{\prime})\phi_{\pm}$,
and integrating ${\bf r}^{\prime}$ over all $y$ and a narrow interval of width $a$
around $x$, one obtains two equations for the coefficients $\psi_{\pm}$, which may
be expressed as
\begin{eqnarray}
(\phi_+|\vec{\sigma}|\phi_-)_{\bf r}\cdot[-i\hbar v_F \vec{\nabla} \psi_-]
+[-i\hbar v_F (\phi_+|\vec{\sigma}\cdot \vec{\nabla} |\phi_-)_{\bf r}]\psi_-
+[-i\hbar v_F (\phi_+|\vec{\sigma}\cdot \vec{\nabla} |\phi_+)_{\bf r}]\psi_+
&= \varepsilon\psi_+ \nonumber\\
(\phi_-|\vec{\sigma}|\phi_+)_{\bf r}\cdot[-i\hbar v_F \vec{\nabla} \psi_+]
+[-i\hbar v_F (\phi_-|\vec{\sigma}\cdot \vec{\nabla} |\phi_+)_{\bf r}]\psi_+
+[-i\hbar v_F (\phi_-|\vec{\sigma}\cdot \vec{\nabla} |\phi_-)_{\bf r}]\psi_-
&= \varepsilon\psi_-. \nonumber\\
\end{eqnarray}
In deriving this expression, we have used the fact that
$(\phi_{\pm}|\sigma_{x,y}|\phi_{\pm})_{\bf r}=0$.

We next demonstrate
that $(\phi_-|\vec{\sigma}\cdot \vec{\nabla} |\phi_-)_{\bf r}=0$.  Writing
\begin{eqnarray}
\vec{\phi}_-[{\bf r}^{\prime},x] \equiv
\left(
\begin{array}{c}
\phi^A_{-}[x^{\prime};V_0(x)] \\
\phi^B_{-}[x^{\prime};V_0(x)]
\end{array}
\right)
e^{ik_y^*[V_0(x)]y^{\prime}},
\end{eqnarray}
we obtain explicitly
\begin{eqnarray}
(\phi_-|\vec{\sigma}\cdot \vec{\nabla} |\phi_-)_{\bf r} =
\int_{x-a/2}^{x+a/2} dx^{\prime} \int_{-\infty}^{\infty} g(y-y^{\prime})
\quad\quad\quad\quad\quad\quad\quad\quad\quad\quad\quad\quad\quad\quad\quad\quad \nonumber \\
\times \Biggl\{
\phi_-^{A}[x^{\prime};V_0(x)]^* \left[ \partial_x + iy^{\prime}(\partial_xk_y^*) \right]
\phi_-^B[x^{\prime};V_0(x)]
+\phi_-^{B}[x^{\prime};V_0(x)]^* \left[ \partial_x + iy^{\prime}(\partial_xk_y^*) \right]
\phi_-^A[x^{\prime};V_0(x)]
\Biggr\}.\nonumber\\
\label{eq2}
\end{eqnarray}
The terms involving $\partial_xk_y^*$ are proportional to
$\int_{x-a/2}^{x+a/2} dx^{\prime} \vec{\phi}_-[{\bf r}^{\prime},\;V_0(x)]^{\dag}
\sigma_x \vec{\phi}_-[{\bf r}^{\prime},\;V_0(x)]$, which vanishes identically, as
in Eq. 2 of the text.  What remains of this matrix element can be cast in the form
\begin{eqnarray}
(\phi_-|\vec{\sigma}\cdot \vec{\nabla} |\phi_-)_{\bf r} &=&
\int_{x-a/2}^{x+a/2} dx^{\prime} \vec{\phi}_-[x^{\prime},y^{\prime}=0;V_0(x)]^{\dag}
\sigma_x\partial_x \vec{\phi}_-[x^{\prime},y^{\prime}=0;V_0(x)] \nonumber \\
&=& \int_{x-a/2}^{x+a/2} dx^{\prime} \frac{1}{{\cal N}_-(x)}
\left[
\vec{\phi}_0[x^{\prime};V_0(x)] - \sigma_z S\vec{\phi}_0[x^{\prime};V_0(x)]
\right]^{\dag} \nonumber \\
&\times& \sigma_x\partial_x \left\lbrace
 \frac{1}{{\cal N}_-(x)}
\left[
\vec{\phi}_0[x^{\prime};V_0(x)] - \sigma_z S\vec{\phi}_0[x^{\prime};V_0(x)]
\right]
\right\rbrace ,
\end{eqnarray}
where in the last expression we have dropped the explicit $y^{\prime}$ in the
arguments of the wavefunctions.  Note the normalization factors ${\cal N}_-$
obtain their spatial dependence from $V_0(x)$.  When the derivative operator
acts on $1/{\cal N}_-$, the resulting contribution is again proportional
to $(\phi_-|\sigma_x|\phi_-)_{\bf r}$ and vanishes.  We now have
\begin{eqnarray}
(\phi_-|\vec{\sigma}\cdot \vec{\nabla} |\phi_-)_{\bf r} &=&
\frac{1}{{\cal N}_-(x)^2}\int_{x-a/2}^{x+a/2} dx^{\prime}\left[
\vec{\phi}_0[x^{\prime};V_0(x)] - \sigma_z S\vec{\phi}_0[x^{\prime};V_0(x)]
\right]^{\dag} \nonumber \\
&\times& \sigma_x\partial_x \left\lbrace
\left[
\vec{\phi}_0[x^{\prime};V_0(x)] - \sigma_z S\vec{\phi}_0[x^{\prime};V_0(x)]
\right]
\right\rbrace \nonumber \\
&=&
\frac{1}{{\cal N}_-(x)^2}\int_{x-a/2}^{x+a/2} dx^{\prime}
\Biggl\{
\vec{\phi}_0^{\dag} \sigma_x\partial_x \vec{\phi_0}
-[S\vec{\phi}_0]^{\dag} \sigma_x\partial_x [S\vec{\phi_0}] \nonumber \\
&&\quad\quad\quad\quad
+[S\vec{\phi}_0]^{\dag} \sigma_x\sigma_z\partial_x \vec{\phi_0}
-\vec{\phi}_0^{\dag} \sigma_x\sigma_z\partial_x [S\vec{\phi_0}]
\Biggr\},
\nonumber \\
\end{eqnarray}
where we used $\{\sigma_x,\sigma_z\}=0$ and $\sigma_z^2=1$.  Finally,
noting that $S$ is a Hermitian operator in $x^{\prime}$ and that $S^2=1$
since the wavefunctions are periodic in $x^{\prime}$, it easy to
see that the four terms of the last expression mutually cancel.  We thus
have established $(\phi_-|\vec{\sigma}\cdot \vec{\nabla} |\phi_-)_{\bf r}=0$.
A very similar manipulation leads to
$(\phi_+|\vec{\sigma}\cdot \vec{\nabla} |\phi_+)_{\bf r}=0$ as well.
Note that in reaching these results, the antisymmetry of $\sigma_x$ and
$\sigma_z$ played a crucial role, so that the vanishing of this
matrix element is in part a consequence of the fact that we can define a
chiral operator for this system.  That
$(\phi_{\pm}|\vec{\sigma}\cdot \vec{\nabla} |\phi_{\pm})_{\bf r}$
vanish is significant because had they not, the matrix equation for
$\psi_{\pm}$ would contain diagonal elements, effectively adding a
scalar potential in addition to the vector potential we find
generated by the modulation of $V_0$.

Our equations now reduce to the form
\begin{eqnarray}
-i\hbar v_F
(\phi_+|\sigma_x|\phi_-)_{\bf r}\partial_x \psi_-
-i\hbar v_F
(\phi_+|\sigma_y|\phi_-)_{\bf r}\partial_y \psi_-
+-i\hbar v_F (\phi_+|\vec{\sigma}\cdot \vec{\nabla} |\phi_-)_{\bf r}\psi_-
&=&\varepsilon\psi_+ \nonumber\\
-i\hbar v_F
(\phi_-|\sigma_x|\phi_+)_{\bf r}\partial_x \psi_+
-i\hbar v_F
(\phi_-|\sigma_y|\phi_+)_{\bf r}\partial_y \psi_+
+-i\hbar v_F (\phi_-|\vec{\sigma}\cdot \vec{\nabla} |\phi_+)_{\bf r}\psi_+
&=&\varepsilon\psi_- \nonumber\\
\end{eqnarray}
Using Eqs. 3 and 4 from the text, we can write four of the coefficients as
position-dependent velocities:
\begin{eqnarray}
\hbar v_F(\phi_+|\sigma_x|\phi_-)_{\bf r} &\equiv& iv_x(x), \nonumber\\
\hbar v_F(\phi_-|\sigma_x|\phi_+)_{\bf r} &\equiv& -iv_x(x), \nonumber\\
\hbar v_F(\phi_+|\sigma_y|\phi_-)_{\bf r} &\equiv& v_y(x), \nonumber\\
\hbar v_F(\phi_-|\sigma_y|\phi_+)_{\bf r} &\equiv& v_y(x). \nonumber\\
\label{velocities}
\end{eqnarray}
Note again that the position dependence of these velocities enters through
the $x$ dependence of $V_0(x)$.
With the further definition
\begin{equation}
a_{\pm,\mp} \equiv -i\hbar v_F (\phi_{\pm}|\vec{\sigma}\cdot \vec{\nabla}
|\phi_{\mp})_{\bf r} =  -i\hbar v_F (\phi_{\pm}|\sigma_x \partial_x
|\phi_{\mp})_{\bf r},
\end{equation}
where the second equality is a consequence of the fact that $V_0$ depends on
$x$ but not on $y$, we arrive at Eq. 5 of the text:
\begin{eqnarray}
\left(
\begin{array}{cc}
0 & v_x\partial_x - i v_y\partial_y + a_{+-}({\bf r}) \\
-v_x\partial_x - i v_y\partial_y + a_{-+}({\bf r}) &  0
\end{array}
\right)
\left(
\begin{array}{c}
\psi_+ \\
\psi_-
\end{array}
\right)
=
\varepsilon
\left(
\begin{array}{c}
\psi_+ \\
\psi_-
\end{array}
\right).
\nonumber
\label{almost_dirac}
\end{eqnarray}

\end{document}